\documentclass[journal]{IEEEtran}

\usepackage{amsmath,amssymb,amsthm,mathtools,bm}
\usepackage{graphicx}
\usepackage{booktabs}
\usepackage{cite}
\usepackage[hidelinks]{hyperref}
\usepackage{microtype}
\usepackage{placeins}
\usepackage{algorithm}
\usepackage{algpseudocode}
\usepackage{xcolor}

\usepackage{tikz}
\usetikzlibrary{arrows.meta,positioning,fit,calc}

\graphicspath{{figures/}}

\newtheorem{assumption}{Assumption}
\newtheorem{lemma}{Lemma}
\newtheorem{theorem}{Theorem}
\newtheorem{corollary}{Corollary}
\theoremstyle{remark}
\newtheorem{remark}{Remark}

\newcommand{\R}{\mathbb{R}}
\newcommand{\Sone}{\mathbb{S}^{1}}
\newcommand{\one}{\bm{1}}
\newcommand{\col}{\operatorname{col}}
\newcommand{\prox}{\operatorname{prox}}

\newcommand{\norm}[1]{\left\lVert#1\right\rVert}
\newcommand{\abs}[1]{\left\lvert#1\right\rvert}
\newcommand{\zc}{z_{\mathrm c}}
\newcommand{\zstar}{z^{*}}
\newcommand{\qstar}{q^{*}}

\newcommand{\wrap}{\operatorname{wrap}_{\pi}}

\title{Distributed Geometric Optimization and Control of
Rigid Differential-Drive Formations from Relative-Position Measurements}

\author{Thanacha~Choopojcharoen, Erman~Selim,  Ramisha Anjum, and Baris~Fidan
\thanks{Preprint version, November 2026. \emph{(Corresponding author: Thanacha Choopojcharoen.)}}
\thanks{This work was supported by the Natural Sciences and Engineering Research Council of Canada (NSERC) Discovery Grant 116806. E.~Selim also thanks the Scientific and Technological Research Council of T\"urkiye (T\"UB\.ITAK) for its support through the 2219 Postdoctoral Program.}
\thanks{T.~Choopojcharoen, R.~Anjum, and B.~Fidan are with the Department of Mechanical and Mechatronics Engineering, University of Waterloo, Waterloo, ON, Canada (e-mail: tchoopoj@uwaterloo.ca; r3anjum@uwaterloo.ca; fidan@uwaterloo.ca).}
\thanks{E.~Selim is with the Department of Electrical and Electronics Engineering, Ege University, Izmir, T\"urkiye (e-mail: erman.selim@ege.edu.tr).}}

\hypersetup{
  pdftitle={Distributed Geometric Optimization and Control of Rigid Differential-Drive Formations from Relative-Position Measurements},
  pdfauthor={Thanacha Choopojcharoen, Erman Selim, Ramisha Anjum, and Baris Fidan}
}

\begin{document}
\raggedbottom

\maketitle

\begin{abstract} Distance-based formation control fixes the shape of a robot team but leaves its position and orientation unconstrained. We regulate all three for differential-drive robots that have only one absolute-position anchor, absolute headings, body-frame relative-position measurements, and communication with their neighbors. We derive a two-layer estimation--control scheme directly from optimization. The first half of the result is a set of graph-proximal estimators that reconstruct the centroid, the anchor vector, and the centered coordinates. Exact proximal updates reach consensus in finitely many steps, and a finite number of primal--dual iterations leaves a bounded reconstruction error. The controller combines proximal centroid and orientation steps with distance-gradient shape regulation, and a finite rotation removes the orientation drift caused by the shape step. A discrete Lyapunov analysis gives local exponential convergence once the estimators have converged. We validate the method in ROS~2/Gazebo and in experiments with five physical robots, where the centroid and orientation errors in the final reference period stay below $1.1$~cm and $0.019$~rad. \end{abstract}

\begin{IEEEkeywords}
Distributed control, multi-agent systems,  primal--dual optimization, differential-drive robots.
\end{IEEEkeywords}

\IEEEpeerreviewmaketitle

\section{Introduction}\label{sec:intro}
Teams of mobile robots are increasingly used for cooperative surveillance \cite{LiuEtAl2021}, spacecraft formations \cite{CuiEtAl2024}, and collaborative transport \cite{ReckerEtAl2021,WuEtAl2024}. These applications benefit from distributed sensing, payload sharing, and redundancy \cite{RenEtAl2020,HuEtAl2021}, but the robots must move as a coherent formation: they must preserve a prescribed shape, track a desired position, and maintain a desired orientation. Meeting all three objectives with only local sensing and sparse communication matters most when reliable global localization is unavailable to most agents.

Distance-based rigid formation control stabilizes inter-agent distances but leaves the two translational and one rotational rigid-body degrees of freedom free \cite{AndersonEtAl2008,KrickEtAl2009}. Rigid motions have been generated through distance mismatches \cite{deMarinaEtAl2016}, and other work has prescribed the formation orientation \cite{SunEtAl2017}, combined formation and trajectory tracking in local coordinates \cite{YangEtAl2018}, or added transport and safety objectives \cite{HerguedasEtAl2022}. More recent work replaces distance constraints with other geometric ones. A leader--follower scheme steers agents to a linear transformation of a nominal configuration \cite{ZhangEtAl2025}, and an angle-constrained law for double-integrator agents rejects trigonometric-polynomial disturbances \cite{PengEtAl2025}. These results provide parts of formation motion, but none of them simultaneously regulates shape, centroid, and formation orientation for a fleet of differential-drive agents that use relative-position measurements.

Our previous work addressed simultaneous centroid estimation and formation tracking under this one-anchor information structure \cite{ChoopojcharoenEtAl2025}. Related foundations are dynamic average consensus \cite{FreemanYangLynch2006,KiaEtAl2019} and finite-time continuous-time trackers for bounded driving signals \cite{ChenCaoRen2012,CaoRenMeng2010}. Centroid and common-frame reconstruction from relative measurements has also been studied \cite{AraguesEtAl2012,FranceschelliGasparri2014}, as has centroid-based collective-motion control \cite{HabibiEtAl2015,SunEtAl2022}, and simultaneous localization and formation control from local relative information is a closely related line of work \cite{GuoEtAl2020,FangEtAl2025}. For nonholonomic vehicles, point-agent coordination laws can be transferred through heading-dependent or near-identity transformations \cite{ZhaoEtAl2018,KhaledyanEtAl2020}. Multirobot platforms highlight the practical importance of sampled communication, local computation, and distributed implementation \cite{WilsonEtAl2020}. Our problem draws on all of these, but it must regulate all three rigid-formation degrees of freedom with only one absolute position available.

Discrete-time dynamic average consensus handles sampled communication directly and characterizes the tradeoff between tracking error and communication \cite{ZhuMartinez2010,MontijanoEtAl2014,FranceschelliGasparri2019}. Distributed subgradient and operator-splitting methods provide local optimization over communication graphs \cite{NedicOzdaglar2009,BoydEtAl2011}. Proximal methods take implicit steps on nonsmooth objectives \cite{ParikhBoyd2014}, and primal--dual splitting lets graph-separable problems be solved through local node and edge updates \cite{ChambollePock2011}. Our sampled-data design builds on these tools. We do not discretize the nonsmooth continuous-time estimator--controller explicitly, because a fixed-step implementation need not preserve its ideal sliding behavior. Instead, we discretize its underlying optimization structure, which leads to an architecture that combines exact-penalty proximal reconstruction with discrete geometric control of the centroid, intrinsic orientation, and shape coordinates.

The contributions are: \begin{itemize} \item \emph{Direct sampled geometric reconstruction}: the centroid, anchor vector, and centered coordinates are reconstructed directly from one absolute-position anchor and neighbor relative-position measurements, without reconstructing individual absolute positions. Exact proximal solves give finite-step reconstruction, and finite-round primal--dual solves admit explicit error bounds.

\item \emph{Discrete pose--shape decoupling}: separate discrete updates regulate the centroid, the intrinsic orientation, and the distance-based shape. A finite rotation exactly removes the orientation drift that the shape step induces, which gives local exponential convergence once the reconstruction is exact.

\item \emph{Distributed finite-computation realization}: the estimator runs as neighbor-only primal--dual iterations, and the commanded controlled-point increments are realized exactly under zero-order hold on differential-drive robots.  \end{itemize}

Sections~\ref{sec:problem}--\ref{sec:analysis} develop the formulation, optimization structure, algorithm, and convergence analysis. Section~\ref{sec:validation} presents the Gazebo validation, the experimental platform and implementation, and the hardware experiments, and Section~\ref{sec:conclusion} concludes the paper.

\section{Problem Formulation}\label{sec:problem}
\subsection{Notation}
$I_m$ is the $m\times m$ identity matrix, $\one_N$ the $N$-vector of ones, and $e_N$ the $N$th canonical vector; $\otimes$ is the Kronecker product, and $\col(x_1,\ldots,x_N)$ stacks its arguments into one column. $\norm\cdot$ is the Euclidean norm. For a stacked edge vector $y=\col(y_1,\ldots,y_M)$ with $y_e\in\R^m$, $\norm y_{2,1}=\sum_e\norm{y_e}$ is the sum of the edge-block norms. For a convex function $f$, $\prox_f(u)=\arg\min_x f(x)+\frac12\norm{x-u}^2$ is its proximal map. $\Sone$ is the set of planar unit vectors (headings), and $\wrap(\cdot)$ maps an angle to $(-\pi,\pi]$. A heading $q\in\Sone$ and an angle $\theta$ both define a rotation matrix: $R(q)=[q\;Sq]$, and $R(\theta)=R([\cos\theta\;\sin\theta]^{\mathsf T})$, where $S$ is the $90^\circ$ rotation in \eqref{eq:S-R}. $\lambda_2(L)$ is the second-smallest eigenvalue of a graph Laplacian $L$; for a connected graph it is positive and sets the rate of the slowest consensus mode.

\subsection{Sampled controlled-point model}
Consider $N\ge3$ differential-drive vehicles. Vehicle $i$ has axle-center position $p_i\in\R^2$, absolute heading $q_i\in\Sone$, longitudinal input $v_i$, and angular input $\omega_i$. Let
\begin{equation}
 S=\begin{bmatrix}0&-1\\1&0\end{bmatrix},\qquad
 R(q_i)=[q_i\;Sq_i]\in SO(2).
 \label{eq:S-R}
\end{equation}
For look-ahead distance $\ell>0$, define the controlled point $z_i=p_i+\ell q_i$. The outer algorithm is sampled at $t_k=kT_s$ and first analyzed through the controlled-point increment model
\begin{equation}
 z_i[k+1]=z_i[k]+\Delta z_i[k].
 \label{eq:sampled-point}
\end{equation}
The centroid, centered coordinate, anchor vector, and intrinsic orientation are $\zc[k]=\frac1N\sum_i z_i[k]$, $\xi_i[k]=z_i[k]-\zc[k]$, \begin{equation}
 r[k]=z_N[k]-\zc[k]=\rho_F[k]q_F[k],
 \qquad q_F[k]\in\Sone.
 \label{eq:anchor-orientation}
\end{equation}
The desired centroid and orientation are $\zstar[k]\in\R^2$ and $\qstar[k]\in\Sone$. Let $\theta_F$, $\theta^*$ denote their angles and $\phi[k]=\wrap(\theta_F[k]-\theta^*[k])$. 
\begin{remark}[Differential-drive realization]
Under zero-order-held $(v_i,\omega_i)$ over one sample, let $a=\omega_iT_s$. Expressed in the body frame at $t_k$, the exact controlled-point increment is
\begin{equation}
 \Delta z_i^{\rm b}=
 \begin{bmatrix}
 \frac{v_i}{\omega_i}\sin a+\ell(\cos a-1)\\[1mm]
 \frac{v_i}{\omega_i}(1-\cos a)+\ell\sin a
 \end{bmatrix},
 \label{eq:zoh-map}
\end{equation}
with the continuous limit $[v_iT_s\;0]^{\mathsf T}$ at $\omega_i=0$, and $\Delta z_i=R(q_i[k])\Delta z_i^{\rm b}$. The Jacobian of this map at $(v_i,\omega_i)=(0,0)$ is $T_s\operatorname{diag}(1,\ell)$, which is nonsingular for $T_s,\ell>0$. By the inverse-function theorem, sufficiently small desired increments in \eqref{eq:sampled-point} therefore have a unique local zero-order-hold realization. We state the convergence analysis for \eqref{eq:sampled-point} accordingly; the local exact realization on a differential-drive robot is a per-agent inversion problem and needs no additional communication. \end{remark}

The controlled point is the standard device for handling the nonholonomic constraint \cite{ZhaoEtAl2018,KhaledyanEtAl2020}: unlike the axle center, $z_i$ can move in any planar direction.

\subsection{Graph, sensing, and rigidity}
Let $\mathcal G=(\mathcal V,\mathcal E)$ be an undirected communication graph with incidence matrix $B\in\R^{N\times M}$ and Laplacian $L=BB^{\mathsf T}$. Define $\Pi=I_N-\frac1N\one_N\one_N^{\mathsf T}$. For each $j\in\mathcal N_i$, agent $i$ measures only $y_{ij}^{\rm b}[k]=R(q_i[k])^{\mathsf T}(z_j[k]-z_i[k])$. Because $q_i[k]$ is known absolutely, each agent can compute the inertial relative displacement $R(q_i[k])y_{ij}^{\rm b}[k]$ locally. No non-anchor knows any absolute position. On-board inertial measurement units can measure the headings, and local sensors such as vision with fiducial markers, marker-less visual pose estimation, or ultra-wideband ranging can measure the relative positions \cite{TsoukalasEtAl2018,WangEtAl2014}. Neighbors share their headings $q_i,q_j$, so a sensor that measures the relative position of the axle centers also gives the controlled-point offset, because $z_j-z_i=p_j-p_i+\ell(q_j-q_i)$. In the experiments of Section~\ref{sec:testbed}, we generate these measurements from motion-capture poses.

\begin{assumption}[Information structure]\label{ass:info}
Only anchor $N$ has access to its own absolute controlled-point position $z_N[k]$. Each agent knows its own absolute heading, measures relative controlled-point positions only to graph neighbors, communicates only with graph neighbors, and knows its own applied increment $\Delta z_i[k]$. The common reference $(\zstar[k],\qstar[k])$ and sample index are available to all agents.
\end{assumption}

\begin{remark}[Role of the anchor]\label{rem:anchor} The absolute headings let every agent express its relative measurements in a common inertial orientation, so relative measurements alone determine the shape and orientation of the formation. They cannot determine where the formation is, because translating every robot by the same vector leaves all measurements unchanged. The single anchor in Assumption~\ref{ass:info} removes this ambiguity, and it enters the algorithm only through the initialization in Section~\ref{sec:algorithm}. The same anchor also defines the formation orientation through the anchor vector $r$ in \eqref{eq:anchor-orientation}. \end{remark}

A framework is rigid when its shape cannot change unless at least one constrained distance changes. It is infinitesimally rigid when this holds to first order; in the plane, this is equivalent to the rigidity matrix having rank $2N-3$ \cite{AndersonEtAl2008,KrickEtAl2009}. A minimally rigid planar graph has exactly $2N-3$ edges, just enough to fix the shape.

\begin{assumption}[Connectivity and rigidity]\label{ass:graph}
The communication graph is connected and need not be complete. The distance-constraint graph is infinitesimally rigid in the neighborhood of the desired planar framework. The two graphs may coincide; when they do, the common graph is connected and infinitesimally rigid.
\end{assumption}

Let $\mathcal E_f$ denote the set of distance-constrained (shape) edges. For an oriented shape edge $e=(i,j)\in\mathcal E_f$ with desired distance $d_{ij}$, define $e_{ij}(z)=\frac12(\norm{z_i-z_j}^2-d_{ij}^2)$, and the distance potential $V_f(z)=\frac12\sum_{(i,j)\in\mathcal E_f}e_{ij}(z)^2$, $g=\nabla_zV_f=R_g(z)^{\mathsf T}e$, where $e=\col_{(i,j)\in\mathcal E_f}(e_{ij})$ and $R_g(z)$ is the rigidity matrix, whose row for edge $(i,j)$ contains $(z_i-z_j)^{\mathsf T}$ in the columns of agent $i$ and $(z_j-z_i)^{\mathsf T}$ in those of agent $j$. Each $g_i$ is computable from the relative measurements on incident formation edges.

\subsection{Problem statement}\label{sec:statement}
Given Assumptions~\ref{ass:info} and~\ref{ass:graph}, design for every agent $i$ a sampled update $\Delta z_i[k]$ that uses only agent $i$'s own heading, its body-frame measurements $y_{ij}^{\rm b}[k]$ to neighbors $j\in\mathcal N_i$, its own applied increments, and messages from its neighbors, such that the following hold.
\begin{enumerate}
 \item \textbf{Formation maintenance:} every constrained inter-agent distance converges to its desired value, $\lim_{k\to\infty}\bigl(\norm{z_i[k]-z_j[k]}-d_{ij}\bigr)=0$, $\forall (i,j)\in\mathcal E_f$.
 \item \textbf{Centroid tracking:} the formation centroid converges to the reference centroid trajectory, $\lim_{k\to\infty}\norm{\zc[k]-\zstar[k]}=0$.
 \item \textbf{Formation-orientation tracking:} the formation orientation defined by the anchor vector in \eqref{eq:anchor-orientation} converges to the reference orientation, $\lim_{k\to\infty}\abs{\phi[k]}=\lim_{k\to\infty}\abs{\wrap\bigl(\theta_F[k]-\theta^*[k]\bigr)}=0$.
\end{enumerate}
Objectives~1--3 fix, respectively, the shape, the two translational degrees of freedom, and the rotational degree of freedom of the planar rigid formation, so together they determine the formation pose completely.

\begin{remark}[Estimation]\label{rem:estimation}
Objectives~2 and~3 refer to the centroid $\zc$ and the orientation $q_F$, which no agent can measure. Each agent $i$ therefore maintains local estimates $\hat c_i[k]$ of $\zc[k]$, $\hat r_i[k]$ of $r[k]$ (hence $\hat q_i[k]$ of $q_F[k]$), and $\hat\xi_i[k]$ of its own centered coordinate $\xi_i[k]$, reconstructed from relative measurements and neighbor messages. The control objectives are met through these estimates, so the analysis in Section~\ref{sec:analysis} first establishes $\hat c_i\to\zc$, $\hat q_i\to q_F$, and $\hat\xi_i\to\xi_i$ and then the three objectives above.
\end{remark}

\section{Optimization Structure}\label{sec:optimization}
\subsection{Geometric tri-objective}
The three objectives of Section~\ref{sec:statement} act on separate parts of the formation geometry, which decomposes naturally into
\begin{equation}
 (\zc,\theta_F,\sigma)\in
 \R^2\times\Sone\times(\R^{2N}/SE(2)),
 \label{eq:product-geometry}
\end{equation}
where $\sigma$ denotes shape modulo rigid translation and rotation. Define $V_c=\frac12\norm{\zc-\zstar}^2$, $V_q=1-q_F^{\mathsf T}\qstar=1-\cos\phi$, $V_f=\frac12\norm e^2$.
Objectives~1, 2, and~3 of Section~\ref{sec:statement} hold exactly when $V_f$, $V_c$, and $V_q$ vanish, respectively. We obtain the outer controller by discrete descent in these three coordinates, not by a Euclidean gradient of $V_c+V_q+V_f$ with respect to the stacked positions.

The distance gradient is orthogonal to rigid translations and to infinitesimal rigid rotation: $(\one_N\otimes a)^{\mathsf T}g=0$, $\tau(\xi)^{\mathsf T}g=0$, where $\tau(\xi)=\col(S\xi_1,\ldots,S\xi_N)$. A shape gradient step therefore changes the quotient-space shape, but it can also turn the anchor vector $\xi_N$. The discrete horizontal-lift construction in Section~IV removes exactly this coordinate artifact.

\subsection{Continuous-time exact penalty and its discrete proximal map}
The nonsmooth continuous estimator, interpreted as a Filippov differential inclusion in the usual sense \cite{Cortes2008}, has the generic disagreement form
\begin{equation}
 \dot\eta\in d(t)-\alpha( L\otimes I_m)\eta
 -\beta(B\otimes I_m)\operatorname{Sgn}((B^{\mathsf T}\otimes I_m)\eta).
 \label{eq:continuous-subgradient}
\end{equation}
It is the subgradient flow of
\begin{equation}
 \Phi_t(\eta)=
 \frac\alpha2\eta^{\mathsf T}(L\otimes I_m)\eta
 +\beta\norm{(B^{\mathsf T}\otimes I_m)\eta}_{2,1}
 -d(t)^{\mathsf T}\eta.
 \label{eq:continuous-objective}
\end{equation}
For $\eta$ in the disagreement subspace and $\norm d\le\Delta$,
\begin{equation}
 \Phi_t(\eta)-\Phi_t(0)
 \ge \frac{\alpha\lambda_2(L)}2\norm\eta^2
 +\mu\norm\eta,
 \label{eq:sharpness}
\end{equation}
where $\mu:=\beta\sqrt{\lambda_2(L)}-\Delta$. Hence $\mu>0$ makes zero disagreement the unique minimizer even when the local driving signals differ between agents. The discrete design keeps this exact-penalty objective and replaces the subgradient flow by its proximal map.

\section{Distributed Discrete-Time Algorithm}\label{sec:algorithm}
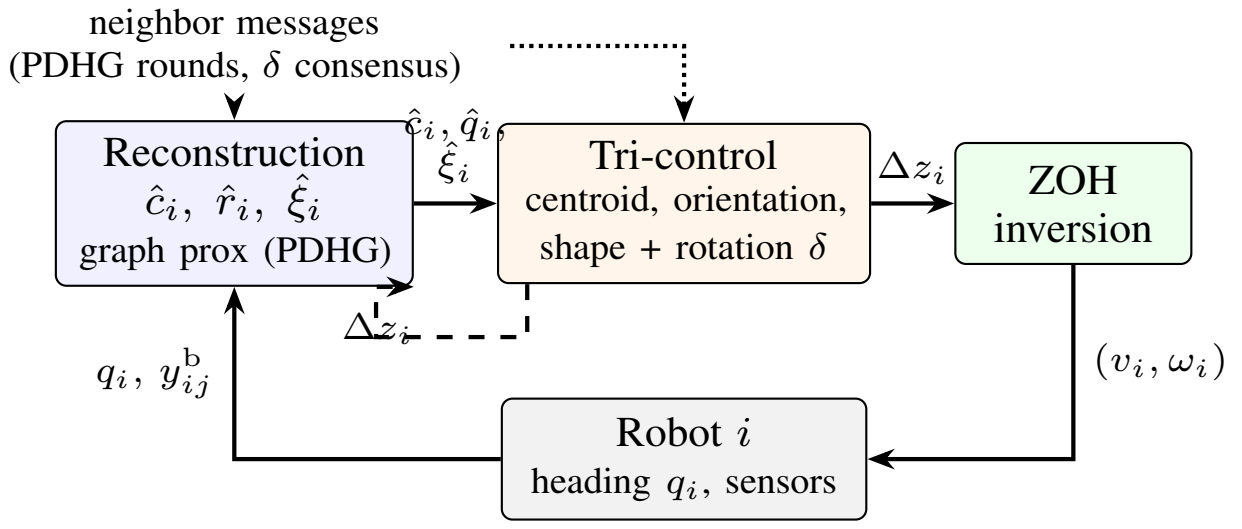
\begin{figure}[!tbp]
 \centering
 \begin{tikzpicture}[
   blk/.style={draw, rounded corners=2pt, align=center, font=\footnotesize, inner sep=3pt, minimum height=8mm},
   arr/.style={-{Stealth[length=1.8mm]}, thick},
   lab/.style={font=\scriptsize, align=center}]
  \node[blk, fill=blue!6, text width=2.15cm] (rec) {Reconstruction\\ $\hat c_i,\ \hat r_i,\ \hat\xi_i$\\ \scriptsize graph prox (PDHG)};
  \node[blk, fill=orange!8, text width=2.25cm, right=0.55cm of rec] (ctl) {Tri-control\\ \scriptsize centroid, orientation,\\ \scriptsize shape + rotation $\delta$};
  \node[blk, fill=green!7, text width=1.35cm, right=0.55cm of ctl] (zoh) {ZOH\\ inversion};
  \node[blk, fill=gray!10, text width=2.2cm, below=0.75cm of ctl] (robot) {Robot $i$\\ \scriptsize heading $q_i$, sensors};
  \draw[arr] (rec) -- node[lab, above]{$\hat c_i,\hat q_i,$\\$\hat\xi_i$} (ctl);
  \draw[arr] (ctl) -- node[lab, above]{$\Delta z_i$} (zoh);
  \draw[arr] (zoh) |- node[lab, pos=0.25, right]{$(v_i,\omega_i)$} (robot);
  \draw[arr] (robot) -| node[lab, pos=0.75, left]{$q_i,\,y^{\rm b}_{ij}$} (rec);
  \draw[arr, dashed] (ctl.south west) ++(0.2,0) |- ++(-1.0,-0.35) |- (rec.south east) ;
  \node[lab, below=0.05cm of rec.south, xshift=0.95cm] {\scriptsize $\Delta z_i$};
  \node[lab, above=0.12cm of rec, text width=3.4cm] (nb) {neighbor messages\\ (PDHG rounds, $\delta$ consensus)};
  \draw[arr, densely dotted] (nb.east) -| (ctl.north);
  \draw[arr, densely dotted] (nb.south) -- (rec.north);
 \end{tikzpicture}%
\caption{Per-agent architecture of the reconstruction, control, and differential-drive realization layers.}

 \label{fig:architecture}
\end{figure}

The algorithm has two layers, which every agent runs at each sample (Fig.~\ref{fig:architecture}). The \emph{reconstruction layer} maintains three local estimates per agent: the centroid estimate $\hat c_i$, the anchor-vector estimate $\hat r_i$ (whose direction $\hat q_i$ estimates the formation orientation), and the centered-coordinate estimate $\hat\xi_i$. Each estimate is updated in two steps. The agent first predicts how the estimate changes because of its own applied motion, and the network then solves a small optimization problem that pulls neighboring estimates together without changing their average. The \emph{control layer} uses these estimates to compute the next increment $\Delta z_i$ from one step for each of the three objectives. Both layers use only local measurements and messages from neighbors.

\subsection{Generic graph proximal tracker}
Let $x_i[k]\in\R^m$ be local estimates and let $f_i[k]$ be local driving signals. Define the predictor
\begin{equation}
 p_i[k]=x_i[k]+T_sf_i[k].
 \label{eq:predictor}
\end{equation}
One exact graph-proximal update is
\begin{equation}
 \begin{aligned}
 x[k+1]=\arg\min_x\;&\frac{1}{2T_s}\sum_i\norm{x_i-p_i[k]}^2\\
 &+\frac\alpha2\sum_{(i,j)\in\mathcal E}\norm{x_i-x_j}^2\\
 &+\beta\sum_{(i,j)\in\mathcal E}\norm{x_i-x_j}.
 \end{aligned}
 \label{eq:generic-prox}
\end{equation}
The first term keeps each estimate close to its own prediction, the quadratic edge term pulls neighboring estimates together smoothly, and the nonsmooth edge term pulls them together with a force of constant magnitude. This constant force is what allows disagreement to vanish exactly rather than only asymptotically. The update is implicit: it is the discrete counterpart of the nonsmooth consensus flow \eqref{eq:continuous-subgradient} evaluated at the end of the sample instead of at its start, so it does not chatter. All coupling terms are edge-local. Summing the first-order optimality conditions gives the exact average identity
\begin{equation}
 \bar x[k+1]=\bar x[k]+T_s\bar f[k].
 \label{eq:average-preservation}
\end{equation}

\begin{lemma}[Finite-step exact consensus]\label{lem:finite-prox}
Suppose $\norm{(\Pi\otimes I_m)f[k]}\le\Delta$ for all $k$ and
\begin{equation}
 \beta>\frac{\Delta}{\sqrt{\lambda_2(L)}}.
 \label{eq:prox-beta-condition}
\end{equation}
If the desired moving quantity has average dynamics consistent with \eqref{eq:average-preservation}, then the disagreement $\eta[k]=(\Pi\otimes I_m)x[k]$ generated by \eqref{eq:generic-prox} reaches zero after finitely many samples. A sufficient bound is
\begin{equation}
 K\le
 \left\lceil\frac{\norm{\eta[0]}^2}{T_s^2\mu^2}\right\rceil,
 \qquad \mu=\beta\sqrt{\lambda_2(L)}-\Delta.
 \label{eq:finite-step-bound}
\end{equation}
Once zero is reached, it is invariant.
\end{lemma}

\begin{proof}
After separating the average mode, \eqref{eq:generic-prox} is $\eta[k+1]=\prox_{T_s\Phi_k}(\eta[k])$, where $\Phi_k$ has the form \eqref{eq:continuous-objective} and satisfies the uniform sharpness inequality \eqref{eq:sharpness}. Such weak-sharp minima are a standard mechanism behind finite termination of proximal methods \cite{BurkeFerris1993,MatsushitaXu2012}. Since $0$ is a minimizer of every $\Phi_k$, firm nonexpansiveness of the proximal map gives $\norm{\eta[k+1]}^2 \le\norm{\eta[k]}^2-\norm{\eta[k]-\eta[k+1]}^2$. If $\eta[k+1]\ne0$, the optimality condition gives $s[k+1]=\frac{\eta[k]-\eta[k+1]}{T_s}\in\partial\Phi_k(\eta[k+1])$. Convexity and \eqref{eq:sharpness} imply $\norm{s[k+1]}\ge\mu$, so the squared norm decreases by at least $T_s^2\mu^2$ until it reaches zero. This yields \eqref{eq:finite-step-bound}. If $\eta[k]=0$, then $\prox_{T_s\Phi_k}(0)=0$ because zero remains the unique minimizer.
\end{proof}

\subsection{Average-matching initialization}
Let $e_N$ be the $N$th canonical vector. Precompute the unique zero-mean vector $h\in\R^N$ satisfying $Lh=\frac1N\one_N-e_N$, $\one_N^{\mathsf T}h=0$. Each agent forms $\ell_i[k]=\sum_{j\in\mathcal N_i}(z_i-z_j)=-\sum_{j\in\mathcal N_i}R(q_i)y_{ij}^{\rm b}$.
Initialize
\begin{align}
 \hat c_i[0]&=Nh_i\ell_i[0]+N\delta_{iN}z_N[0],\label{eq:c-init-d}\\
 \hat r_i[0]&=-Nh_i\ell_i[0],\label{eq:r-init-d}\\
 \hat\xi_i[0]&=0.\label{eq:xi-init-d}
\end{align}
Then
\begin{align}
 \frac1N\sum_i\hat c_i[0]&=\zc[0],\qquad
 \frac1N\sum_i\hat r_i[0]=r[0],\nonumber\\
 \sum_i\hat\xi_i[0]&=\sum_i\xi_i[0]=0.
 \label{eq:initial-average-match}
\end{align}
Only the anchor uses an absolute position in \eqref{eq:c-init-d}.

\subsection{Centroid and orientation reconstruction}
After applying $\Delta z_i[k]$, define $f_i^c[k]=\frac{\Delta z_i[k]}{T_s}$. The centroid estimate $\hat c[k+1]$ is obtained from \eqref{eq:generic-prox} with $x=\hat c$, $f=f^c$, and gains $(\alpha_c,\beta_c)$. Since $\zc[k+1]-\zc[k]=\frac1N\sum_i\Delta z_i[k]$, \eqref{eq:average-preservation} and \eqref{eq:initial-average-match} then ensure that the estimate average equals the true centroid at every sample.

For the anchor vector, use $f_i^r[k]=\frac{(N\delta_{iN}-1)\Delta z_i[k]}{T_s}$. Its average is $(r[k+1]-r[k])/T_s$, so applying \eqref{eq:generic-prox} with gains $(\alpha_r,\beta_r)$ gives a finite-step exact estimate of $r[k]$. Define
\begin{equation}
 \hat q_i[k]=\frac{\hat r_i[k]}{\norm{\hat r_i[k]}}
 \label{eq:qhat-d}
\end{equation}
whenever $\hat r_i[k]\ne0$.

\begin{corollary}[Centroid and orientation consensus]\label{cor:c-r}
If $\rho_F[k]\ge\rho_{\min}>0$ and the exact-penalty conditions of Lemma~\ref{lem:finite-prox} hold for $f^c$ and $f^r$, then there exist finite $K_c,K_r$ such that $\hat c_i[k]=\zc[k]$ for $k\ge K_c$, and $\hat r_i[k]=r[k]$, $\hat q_i[k]=q_F[k]$ for $k\ge K_r$, for every agent. Hence the centroid and intrinsic-orientation estimates reach exact network consensus.
\end{corollary}

\subsection{Direct centered-coordinate reconstruction}
A non-anchor cannot form $z_i-\hat c_i$ because it does not know absolute $z_i$. For an oriented communication edge $e=(i,j)$, the measured inertial offset at sample $k+1$ is
\begin{equation}
 d_e[k+1]=z_i[k+1]-z_j[k+1]
 =-R(q_i[k+1])y_{ij}^{\rm b}[k+1].
 \label{eq:edge-offset}
\end{equation}
Let $d[k+1]=\col_e(d_e[k+1])$. Define
\begin{equation}
 p_i^{\xi}[k]=\hat\xi_i[k]+\Delta z_i[k]
 -(\hat c_i[k+1]-\hat c_i[k]).
 \label{eq:xi-predictor}
\end{equation}
Then compute
\begin{equation}
 \begin{split}
 \hat\xi[k+1]=\arg\min_x\;&\frac{1}{2T_s}\sum_i\norm{x_i-p_i^{\xi}[k]}^2\\
 &+\frac{\alpha_\xi}{2}\norm{(B^{\mathsf T}\otimes I_2)x-d[k+1]}^2\\
 &+\beta_\xi\norm{(B^{\mathsf T}\otimes I_2)x-d[k+1]}_{2,1}.
 \end{split}
 \label{eq:xi-prox}
\end{equation}
In error coordinates $\tilde\xi=\hat\xi-\xi$, \eqref{eq:xi-prox} is again a proximal exact-penalty recursion. Its zero-average property follows from \eqref{eq:xi-predictor}, because the average increment of the centroid estimator equals the true centroid increment. Lemma~\ref{lem:finite-prox} therefore applies to $\tilde\xi$ under a uniform bound on the induced disagreement forcing.

\subsection{Discrete optimization-derived tri-control}
At sample $k$, each agent uses its current estimates to compute the next desired centroid coordinate. Let $\gamma_c=\frac{1}{1+T_sk_c}$. The minimizer of
\begin{equation*}
 \min_c\;
 \frac{1}{2T_s}\norm{c-(\hat c_i[k]+\Delta z^*[k])}^2
 +\frac{k_c}{2}\norm{c-\zstar[k+1]}^2,
\end{equation*}
where $\Delta z^*[k]=\zstar[k+1]-\zstar[k]$, is $c_i^d[k+1]=\zstar[k+1] +\gamma_c(\hat c_i[k]-\zstar[k])$. 
For orientation, let $\hat\theta_i[k]=\operatorname{atan2}(\hat q_{i,2},\hat q_{i,1})$ and $\hat\phi_i[k]=\wrap(\hat\theta_i[k]-\theta^*[k])$. The proximal orientation error is
\begin{equation*}
 \phi_i^+[k]=\arg\min_{\varphi\in(-\pi,\pi]}
 \frac{(\varphi-\hat\phi_i[k])^2}{2T_s}
 +k_q(1-\cos\varphi),
\end{equation*}
which satisfies the scalar implicit equation
\begin{equation}
 \phi_i^+[k]+T_sk_q\sin\phi_i^+[k]=\hat\phi_i[k].
 \label{eq:orientation-implicit}
\end{equation}
Thus the desired rigid orientation increment is
\begin{equation}
 \Delta\theta_i^d[k]
 =\theta^*[k+1]-\theta^*[k]+\phi_i^+[k]-\hat\phi_i[k].
 \label{eq:orientation-increment}
\end{equation}

The shape descent is
\begin{equation}
 y_i[k]=\hat\xi_i[k]-T_sk_fg_i[k].
 \label{eq:shape-step}
\end{equation}
The shape step corrects only inter-agent distances, but it can also turn the anchor vector and, with it, the formation orientation that the orientation step controls. To undo this, the whole shape target is rotated back by the angle through which the shape step turned the anchor vector. We call this rotation the horizontal-lift correction because, geometrically, it lifts a pure shape change to a motion that leaves the orientation unchanged. The anchor computes the correction
\begin{equation}
 \delta_s[k]=\wrap\!\left(
 \hat\theta_N[k]-\operatorname{atan2}(y_{N,2}[k],y_{N,1}[k])
 \right).
 \label{eq:delta-shape}
\end{equation}
For small $T_s$, $\delta_s/T_s$ tends to the continuous compensation $k_f(Sq_F)^{\mathsf T}g_N/\rho_F$, whereas \eqref{eq:delta-shape} exactly removes the change in anchor-vector direction caused by the finite shape step.

The scalar $\delta_s[k]$ is distributed by solving the pinned graph problem
\begin{equation}
 \min_{\delta_1,\ldots,\delta_N}
 \frac12\sum_{(i,j)\in\mathcal E}(\delta_i-\delta_j)^2,
 \qquad \delta_N=\delta_s[k].
 \label{eq:pinned-delta}
\end{equation}
Connectedness makes its unique minimizer $\delta_i=\delta_s[k]$ for all $i$. Let $\hat\delta_i[k]$ denote the distributed solution. The complete local shape/orientation target is
\begin{equation}
 \xi_i^d[k+1]
 =R\!\left(\hat\delta_i[k]+\Delta\theta_i^d[k]\right)y_i[k].
 \label{eq:xi-desired}
\end{equation}
Finally,
\begin{equation}
 \Delta z_i[k]
 =c_i^d[k+1]-\hat c_i[k]
 +\xi_i^d[k+1]-\hat\xi_i[k].
 \label{eq:discrete-tricontrol}
\end{equation}
Every quantity in \eqref{eq:discrete-tricontrol} is local, measured from a neighbor, or obtained from a neighbor-only distributed optimization subproblem. The first difference in \eqref{eq:discrete-tricontrol} moves the formation along the reference and contracts the centroid error by the factor $\gamma_c$ per sample. The second difference changes only the position of the agent relative to the centroid. It corrects the shape through $y_i$, rotates the whole formation rigidly toward $\theta^*$ through $\Delta\theta_i^d$, and cancels the effect of the shape step on the orientation through $\hat\delta_i$. Rigid rotations preserve distances and $\sum_i\xi_i^d=0$ preserves the centroid, so the three steps act on separate coordinates of \eqref{eq:product-geometry} and achieve the three objectives jointly without competing.

\subsection{Neighbor-only primal--dual realization}
Problems \eqref{eq:generic-prox} and \eqref{eq:xi-prox} have the form
\begin{equation*}
 \min_x F_k(x)+G(\mathcal B^{\mathsf T}x-d),
 \qquad \mathcal B=B\otimes I_m,
\end{equation*}
with a separable strongly convex node term $F_k$ and separable edge term $G(y)=\frac\alpha2\norm y^2+\beta\norm y_{2,1}$. A primal--dual hybrid-gradient iteration \cite{ChambollePock2011} is
\begin{align}
 \lambda^{\ell+1}&=\prox_{\sigma G^*}
 \left(\lambda^\ell+\sigma(\mathcal B^{\mathsf T}\bar x^\ell-d)\right),\label{eq:pdhg-dual}\\
 x^{\ell+1}&=\prox_{\tau F_k}
 \left(x^\ell-\tau\mathcal B\lambda^{\ell+1}\right),\label{eq:pdhg-primal}\\
 \bar x^{\ell+1}&=x^{\ell+1}+\vartheta(x^{\ell+1}-x^\ell),\nonumber
\end{align}
with $\tau\sigma\norm{\mathcal B}^2<1$. Each edge update in \eqref{eq:pdhg-dual} uses only its two endpoint states and the local edge measurement, and each node update in \eqref{eq:pdhg-primal} uses only its predictor and the dual messages of its incident edges. The realization is therefore distributed and needs neither all-to-all communication nor a central matrix inversion.

For every outer sample $k$, initializing the primal inner iteration at the predictor,
\begin{equation}
 x^{0,k}=p[k],
 \label{eq:pdhg-primal-init}
\end{equation}
with the dual variable warm-started from the preceding outer sample, preserves the network-average identity of the proximal subproblem for any finite number of inner iterations. To see this, consider the quadratic node term $F_k(x)=\frac{1}{2T_s}\norm{x-p[k]}^2$, whose proximal map is
\begin{equation}
 \prox_{\tau F_k}(u)
 =\frac{u+(\tau/T_s)p[k]}{1+\tau/T_s}.
 \label{eq:F-prox-closed}
\end{equation}
Since $(\one_N^{\mathsf T}\otimes I_m)\mathcal B=0$, premultiplying \eqref{eq:pdhg-primal} by $(\one_N^{\mathsf T}\otimes I_m)/N$ and using \eqref{eq:F-prox-closed} gives, for the network average $\bar x^{\ell,k}$ of the primal iterate,
\begin{equation}
 \bar x^{\ell+1,k}-\bar p[k]=\frac{\bar x^{\ell,k}-\bar p[k]}{1+\tau/T_s}.
 \label{eq:pdhg-average-recursion}
\end{equation}
Under \eqref{eq:pdhg-primal-init}, \eqref{eq:pdhg-average-recursion} shows inductively that
\begin{equation}
 \frac1N(\one_N^{\mathsf T}\otimes I_m)x^{\ell,k}=\bar p[k],
 \qquad \ell=0,1,\ldots .
 \label{eq:finite-pdhg-average}
\end{equation}
Terminating PDHG after a finite number of rounds therefore introduces only disagreement-mode error and does not perturb the average required by \eqref{eq:average-preservation}.

The implementations used in Sections~\ref{sec:gazebo}--\ref{sec:physical} instead initialize the primal iteration at the current estimate, $x^{0,k}=x[k]$. Since $\bar x[k]-\bar p[k]=-T_s\bar f[k]$, \eqref{eq:pdhg-average-recursion} then gives, after $J$ rounds,
\begin{equation}
 \norm{\bar x^{J,k}-\bar p[k]}=\Bigl(1+\frac{\tau}{T_s}\Bigr)^{-J}T_s\norm{\bar f[k]}.
 \label{eq:pdhg-average-error}
\end{equation}
Because $1+\tau/T_s>1$, this error decays geometrically with the number of inner rounds and vanishes as $J\to\infty$. It is proportional to the average motion per sample, $T_s\norm{\bar f[k]}$, and over $K$ samples it accumulates to at most $\sum_{k<K}(1+\tau/T_s)^{-J}T_s\norm{\bar f[k]}$. It is therefore negligible for a moderate number of rounds, and \eqref{eq:finite-pdhg-average} and the conclusions drawn from it hold up to this term.

\begin{remark}[Exact versus finite inner solves]\label{rem:inner} The exact proximal recursion and its finite-round distributed realization converge differently. The exact minimizer keeps the finite-step property of Lemma~\ref{lem:finite-prox}. A finite number of PDHG rounds generally does not compute that minimizer exactly, so it need not produce finite-step exact consensus. The analysis below quantifies this implementation error: a bounded inner-solve error gives an explicit practical-consensus tube, and a vanishing inner-solve error recovers exact asymptotic convergence. The analysis assumes the predictor initialization \eqref{eq:pdhg-primal-init}. With the current-estimate initialization used in the implementation, the inner-solve error also contains the average component \eqref{eq:pdhg-average-error}, which is negligible. The exact-proximal closed-loop theorem is kept separately as the ideal limiting case. \end{remark}

\subsection{Per-agent estimation and control laws}\label{sec:per-agent}
For implementation, the stacked operations above reduce to explicit updates that each agent performs with its own data and messages from its neighbors. Let $\mathcal N_i^f$ be the set of shape-edge neighbors of agent $i$, and let $b_{ie}=1$ if agent $i$ is the first endpoint of the oriented edge $e$, $b_{ie}=-1$ if it is the second, and $b_{ie}=0$ otherwise. Let $\wrap(\cdot)$ map an angle to $(-\pi,\pi]$ and $\operatorname{wrap}_{\pi/2}(\cdot)$ map it to $[-\pi/2,\pi/2)$ modulo $\pi$.

\emph{Control law.} At sample $k$, agent $i$ computes its increment from its heading $q_i[k]$, its measurements $y^{\rm b}_{ij}[k]$, and its estimates $\hat c_i[k]$, $\hat r_i[k]$, $\hat\xi_i[k]$:
\begin{align}
 g_i&=\sum_{j\in\mathcal N_i^f}\tfrac12\bigl(\norm{s_{ij}}^2-d_{ij}^2\bigr)s_{ij},
 \ s_{ij}=-R(q_i[k])y^{\rm b}_{ij}[k],\label{eq:cl-gradient}\\
 c_i^d&=\zstar[k+1]+\gamma_c\bigl(\hat c_i[k]-\zstar[k]\bigr),\label{eq:cl-centroid}\\
 \hat\phi_i&=\wrap\bigl(\operatorname{atan2}(\hat r_{i,2}[k],\hat r_{i,1}[k])-\theta^*[k]\bigr),\label{eq:cl-phihat}\\
 \phi_i^+&:\ \phi_i^++T_sk_q\sin\phi_i^+=\hat\phi_i,\label{eq:cl-phiplus}\\
 \Delta\theta_i^d&=\theta^*[k+1]-\theta^*[k]+\phi_i^+-\hat\phi_i,\label{eq:cl-dtheta}\\
 y_i&=\hat\xi_i[k]-T_sk_fg_i,\label{eq:cl-shape}\\
 \Delta z_i[k]&=c_i^d-\hat c_i[k]+R\bigl(\hat\delta_i[k]+\Delta\theta_i^d\bigr)y_i-\hat\xi_i[k],\label{eq:cl-increment}\\
 \Delta z_i[k]&\leftarrow\min\Bigl(1,\frac{u_{\max}T_s}{\norm{\Delta z_i[k]}}\Bigr)\Delta z_i[k].\label{eq:cl-saturation}
\end{align}
Newton's method, started at $\phi_i^+=\hat\phi_i$, solves the scalar equation \eqref{eq:cl-phiplus}; its left-hand side is strictly increasing whenever $T_sk_q<1$, which holds for all runs reported here. The rotation angle $\hat\delta_i[k]$ in \eqref{eq:cl-increment} is obtained by $M$ rounds of neighbor-only gradient descent on the pinned problem \eqref{eq:pinned-delta}:
\begin{equation}
 \begin{aligned}
 \delta_N^m&=\delta_s[k]=\wrap\bigl(\operatorname{atan2}(\hat r_{N,2},\hat r_{N,1})-\operatorname{atan2}(y_{N,2},y_{N,1})\bigr),\\
 \delta_i^{m+1}&=\delta_i^m-\gamma_\delta\sum_{j\in\mathcal N_i}\bigl(\delta_i^m-\delta_j^m\bigr),\qquad i\ne N,
 \end{aligned}
 \label{eq:cl-delta}
\end{equation}
for $m=0,\ldots,M-1$, with $\delta_i^0=\hat\delta_i[k-1]$, $\hat\delta_i[k]=\delta_i^M$, and step $\gamma_\delta=\min\{0.16,\,0.95/\max_i\abs{\mathcal N_i}\}$. Finally, the saturated increment \eqref{eq:cl-saturation}
is stored for the estimator and converted into the wheel commands that realize it exactly under zero-order hold, by inverting \eqref{eq:zoh-map}:
\begin{equation}
 \begin{gathered}
 \begin{bmatrix}a_i\\ b_i\end{bmatrix}=R(q_i[k])^{\mathsf T}\Delta z_i[k],\qquad
 \alpha_i=\operatorname{wrap}_{\pi/2}\bigl(\operatorname{atan2}(b_i,\,a_i+2\ell)\bigr),\\
 \omega_i=\frac{2\alpha_i}{T_s},\qquad
 v_i=\frac{\alpha_i\,(a_i\cos\alpha_i+b_i\sin\alpha_i)}{T_s\sin\alpha_i},
 \end{gathered}
 \label{eq:cl-zoh-inverse}
\end{equation}
with $v_i=a_i/T_s$ when $\alpha_i=0$. Substituting \eqref{eq:cl-zoh-inverse} into \eqref{eq:zoh-map} recovers $\Delta z_i[k]$ exactly. Wrapping $\alpha_i$ selects the shorter turn, so that $\abs{\omega_i}T_s\le\pi$, and allows reverse motion ($v_i<0$).

\emph{Estimation law.} After applying the increment, agent $i$ forms the predictors
\begin{equation}
 p_i^c=\hat c_i[k]+\Delta z_i[k],\qquad
 p_i^r=\hat r_i[k]+(N\delta_{iN}-1)\Delta z_i[k].
 \label{eq:el-predictors}
\end{equation}
Each estimate is then updated by $J$ rounds of the per-edge and per-node PDHG updates
\begin{align}
 w_e&=\lambda_e^\ell+\sigma\bigl(\bar x_i^\ell-\bar x_j^\ell-d_e\bigr),\quad e=(i,j)\in\mathcal E,\label{eq:el-edge-trial}\\
 \lambda_e^{\ell+1}&=w_e-\frac{\sigma}{\sigma+\alpha}\max\Bigl(0,1-\frac{\beta}{\norm{w_e}}\Bigr)w_e,\label{eq:el-edge}\\
 x_i^{\ell+1}&=\frac{x_i^\ell-\tau\sum_{e\in\mathcal E}b_{ie}\lambda_e^{\ell+1}+(\tau/T_s)\,p_i}{1+\tau/T_s},\label{eq:el-node}\\
 \bar x_i^{\ell+1}&=x_i^{\ell+1}+\vartheta\bigl(x_i^{\ell+1}-x_i^\ell\bigr),\label{eq:el-relax}
\end{align}
for $\ell=0,\ldots,J-1$, initialized with $x_i^0=\bar x_i^0=x_i[k]$ and with $\lambda_e^0$ equal to the final dual of the preceding sample, and returning $x_i[k+1]=x_i^J$. These updates are applied first with $(x,p,\alpha,\beta,d_e)=(\hat c,p^c,\alpha_c,\beta_c,0)$ and $(\hat r,p^r,\alpha_r,\beta_r,0)$. The centered coordinate then uses the predictor
\begin{equation}
 p_i^\xi=\hat\xi_i[k]+\Delta z_i[k]-\bigl(\hat c_i[k+1]-\hat c_i[k]\bigr)
 \label{eq:el-xi-predictor}
\end{equation}
in \eqref{eq:el-edge-trial}--\eqref{eq:el-relax} with $(x,p,\alpha,\beta)=(\hat\xi,p^\xi,\alpha_\xi,\beta_\xi)$ and the freshly measured edge offsets $d_e=d_e[k+1]=-R(q_i[k+1])y^{\rm b}_{ij}[k+1]$ of \eqref{eq:edge-offset}. Finally, $\hat q_i[k+1]=\hat r_i[k+1]/\norm{\hat r_i[k+1]}$. At $k=0$, the estimates are initialized by \eqref{eq:c-init-d}--\eqref{eq:xi-init-d}. The edge update \eqref{eq:el-edge} is the closed form of $\prox_{\sigma G^*}$ in \eqref{eq:pdhg-dual} for the edge term $G_e(y)=\frac\alpha2\norm y^2+\beta\norm y$, and \eqref{eq:el-node} is \eqref{eq:pdhg-primal} with \eqref{eq:F-prox-closed}. An edge update needs only the relaxed estimates of its two endpoints and the offset of that edge. In every round, agent $i$ therefore sends $\bar x_i^\ell$ to its neighbors, one endpoint of each edge computes $\lambda_e^{\ell+1}$ and returns it to the other, and each agent completes \eqref{eq:el-node} with the duals of its incident edges. The step sizes are $\tau=\sigma=0.88/\sqrt{\lambda_{\max}(L)}$, which satisfies $\tau\sigma\norm{\mathcal B}^2<1$ because $\norm{\mathcal B}^2=\lambda_{\max}(L)$, and $\vartheta=1$.

Algorithm~\ref{alg:agent} summarizes the order of these operations for one agent and one sample.

\begin{algorithm}[!tbp]
 \caption{Agent $i$ at sample $k$}
 \label{alg:agent}
 \small
 \begin{algorithmic}[1]
  \Require $q_i[k]$, $y^{\rm b}_{ij}[k]$, $\hat c_i[k],\hat r_i[k],\hat\xi_i[k]$, $\hat\delta_i[k-1]$, duals $\lambda_e$, reference $\zstar,\theta^*$
  \State compute $\Delta z_i[k]$ by \eqref{eq:cl-gradient}--\eqref{eq:cl-saturation}, with $\hat\delta_i[k]$ from \eqref{eq:cl-delta} \Comment{$M$ neighbor rounds}
  \State apply $(v_i,\omega_i)$ from \eqref{eq:cl-zoh-inverse} for one sample
  \State update $\hat c_i,\hat r_i$ by \eqref{eq:el-predictors}--\eqref{eq:el-relax} \Comment{$J$ neighbor rounds each}
  \State measure $y^{\rm b}_{ij}[k+1]$; update $\hat\xi_i$ by \eqref{eq:el-xi-predictor} and \eqref{eq:el-edge-trial}--\eqref{eq:el-relax}
  \State $\hat q_i[k+1]\gets\hat r_i[k+1]/\norm{\hat r_i[k+1]}$
 \end{algorithmic}
\end{algorithm}

\section{Discrete-Time Convergence Analysis}\label{sec:analysis}
The analysis proceeds in three steps. We first bound the reconstruction error when each proximal problem is solved by a finite number of inner iterations (Section~\ref{sec:inexact}). We then show that exact proximal solves reconstruct the geometry exactly after finitely many samples (Section~\ref{sec:exact}). Finally, we show that, once the geometry is exact, each of the three objectives contracts independently (Sections~\ref{sec:centroid-contraction}--\ref{sec:shape-contraction}).

\subsection{Finite-inner-iteration estimator bound}\label{sec:inexact}
Let $P_k$ denote the exact proximal map of the disagreement recursion in Lemma~\ref{lem:finite-prox}. On the disagreement subspace $\mathcal D=\{\eta:(\one_N^{\mathsf T}\otimes I_m)\eta=0\}$, the quadratic graph term is $m_\Phi$-strongly convex with $m_\Phi=\alpha\lambda_2(L)$. The remaining nonsmooth edge penalty is convex and the local driving term is affine, so the complete function $\Phi_k$ remains $m_\Phi$-strongly convex on $\mathcal D$. Hence its proximal map is contractive:
\begin{equation}
 \norm{P_k(\eta_1)-P_k(\eta_2)}
 \le q\norm{\eta_1-\eta_2},
 \ 
 q=\frac{1}{1+T_s\alpha\lambda_2(L)}<1.
 \label{eq:prox-contraction}
\end{equation}
Under the exact-penalty condition \eqref{eq:prox-beta-condition}, zero disagreement is the unique minimizer of every $\Phi_k$, so $P_k(0)=0$.

Let $x^\star[k+1]$ denote the exact minimizer of the sample-$k$ proximal subproblem and let $x[k+1]$ be the output returned after a finite number $J_k$ of PDHG iterations initialized according to \eqref{eq:pdhg-primal-init}. Define
\begin{equation}
 e_k=x[k+1]-x^\star[k+1],
 \qquad \norm{e_k}\le\varepsilon_k.
 \label{eq:inner-error}
\end{equation}
Equation \eqref{eq:finite-pdhg-average} implies that $e_k\in\mathcal D$.

\begin{theorem}[Inexact graph-proximal tracking]\label{thm:inexact-prox}
Suppose the hypotheses of Lemma~\ref{lem:finite-prox} hold and the finite PDHG realization satisfies \eqref{eq:pdhg-primal-init} and \eqref{eq:inner-error}. Then the network average obeys the exact identity \eqref{eq:average-preservation} for every finite $J_k$, while the disagreement satisfies
\begin{equation}
 \norm{\eta[k]}
 \le q^k\norm{\eta[0]}
 +\sum_{j=0}^{k-1}q^{k-1-j}\varepsilon_j,
 \label{eq:inexact-convolution}
\end{equation}
where $q$ is given by \eqref{eq:prox-contraction}. Consequently:
\begin{enumerate}
 \item if $\varepsilon_k\le\bar\varepsilon$ uniformly, then
 \begin{equation}
  \limsup_{k\to\infty}\norm{\eta[k]}
  \le\frac{\bar\varepsilon}{1-q};
  \label{eq:practical-consensus-tube}
 \end{equation}
 \item if $\varepsilon_k\to0$, then $\eta[k]\to0$;
 \item if $\varepsilon_k\le C_\varepsilon\rho_\varepsilon^k$ for some $0<\rho_\varepsilon<1$, then $\norm{\eta[k]} =O\!\left(\max\{q,\rho_\varepsilon\}^k\right)$. \end{enumerate}
\end{theorem}

\begin{proof}
By \eqref{eq:finite-pdhg-average}, the exact and finite-round solutions have the same network average. In disagreement coordinates, the finite-round update can therefore be written as $\eta[k+1]=P_k(\eta[k])+e_k$. Using $P_k(0)=0$ and \eqref{eq:prox-contraction} gives $\norm{\eta[k+1]} \le q\norm{\eta[k]}+\varepsilon_k$. Iteration yields \eqref{eq:inexact-convolution}. If $\varepsilon_k\le\bar\varepsilon$, summing the geometric series gives \eqref{eq:practical-consensus-tube}. If $\varepsilon_k\to0$, the stable convolution in \eqref{eq:inexact-convolution} tends to zero. For geometrically decaying $\varepsilon_k$, evaluating the convolution gives the stated geometric rate.
\end{proof}

The optimization accuracy itself also bounds the inner error. Let $Q_k(x)=F_k(x)+G(\mathcal B^{\mathsf T}x-d)$ be the proximal objective and suppose the finite inner solve satisfies $Q_k(x[k+1])-Q_k(x^\star[k+1])\le\delta_k$. Because the two points have the same network average, their difference lies in $\mathcal D$, on which $Q_k$ is strongly convex with modulus $m_Q=T_s^{-1}+\alpha\lambda_2(L)$. Therefore
\begin{equation}
 \varepsilon_k
 \le
 \sqrt{\frac{2\delta_k}{T_s^{-1}+\alpha\lambda_2(L)}}.
 \label{eq:objective-to-prox-error}
\end{equation}
Equations \eqref{eq:practical-consensus-tube} and \eqref{eq:objective-to-prox-error} relate a finite inner-solve tolerance directly to the resulting estimator disagreement tube.

\begin{corollary}[Finite-round geometric reconstruction]\label{cor:finite-round-geometry}
Apply Theorem~\ref{thm:inexact-prox} separately to the centroid, anchor-vector, and centered-coordinate proximal problems, with respective contraction factors
 $q_s=1/(1+T_s\alpha_s\lambda_2(L))$ for $s\in\{c,r,\xi\}$. Because the corresponding network averages match the true centroid, anchor vector, and zero-centered coordinate average, respectively, the stacked geometric estimation errors satisfy the bounds \eqref{eq:inexact-convolution} with their respective inner-solve errors $\varepsilon_k^c$, $\varepsilon_k^r$, and $\varepsilon_k^\xi$. In particular, uniformly bounded finite-round errors imply uniformly bounded practical reconstruction errors. If $\rho_F[k]\ge\rho_{\min}>0$ and the anchor-vector estimation error is eventually smaller than $\rho_{\min}/2$, normalization in \eqref{eq:qhat-d} additionally gives $\norm{\hat q_i[k]-q_F[k]}\le \frac{2}{\rho_{\min}}\norm{\hat r_i[k]-r[k]}$,
so the intrinsic-orientation estimate inherits the same practical or asymptotic convergence character as the anchor-vector estimate.
\end{corollary}

\subsection{Exact-proximal estimator convergence}\label{sec:exact}
Let $K_c,K_r,K_\xi$ denote the finite settling indices furnished by Lemma~\ref{lem:finite-prox} for the exact centroid, anchor-vector, and centered-coordinate proximal maps, and define $K_E=\max\{K_c,K_r,K_\xi\}$. For $k\ge K_E$,
\begin{equation}
 \hat c_i[k]=\zc[k],\qquad
 \hat q_i[k]=q_F[k],\qquad
 \hat\xi_i[k]=\xi_i[k]
 \label{eq:exact-geometry-after-KE}
\end{equation}
for every agent. Within each outer sample, the exact solution of \eqref{eq:pinned-delta} also gives $\hat\delta_i[k]=\delta_s[k]$.

\subsection{Centroid contraction}\label{sec:centroid-contraction}
Under \eqref{eq:exact-geometry-after-KE}, averaging \eqref{eq:discrete-tricontrol} and using $\sum_i\xi_i^d=\sum_i\xi_i=0$ gives $\zc[k+1]=\zstar[k+1]+\gamma_c(\zc[k]-\zstar[k])$. Hence the centroid tracking error $e_c[k]=\zc[k]-\zstar[k]$ satisfies
\begin{equation}
 e_c[k+1]=\gamma_c e_c[k],
 \qquad 0<\gamma_c<1
 \label{eq:centroid-contraction}
\end{equation}
for every $T_s>0$ and $k_c>0$.

\subsection{Orientation contraction}
After estimator settling, all agents share the same $\phi[k]$. Equation \eqref{eq:orientation-implicit} becomes $\phi[k+1]+T_sk_q\sin\phi[k+1]=\phi[k]$. If $\abs{\phi[k]}\le\bar\phi<\pi/2$, the map $x\mapsto x+T_sk_q\sin x$ has derivative at least $1+T_sk_q\cos\bar\phi$. Therefore
\begin{equation}
 \abs{\phi[k+1]}
 \le\rho_q\abs{\phi[k]},\qquad
 \rho_q=\frac{1}{1+T_sk_q\cos\bar\phi}<1.
 \label{eq:q-contraction}
\end{equation}
The same result follows from the proximal decrease of $V_q=1-\cos\phi$.

\subsection{Shape descent and discrete horizontal lift}\label{sec:shape-contraction}
For the exact geometry, define the unrotated shape step $\xi^s[k]=\xi[k]-T_sk_fg[k]$. Because $\sum_i g_i=0$, $\xi^s$ remains centered. Assume $\nabla V_f$ is $L_f$-Lipschitz in a rigidity neighborhood $\mathcal U$. The descent lemma gives
\begin{equation*}
 V_f(\xi^s[k])\le V_f(\xi[k])
 -T_sk_f\left(1-\frac{T_sk_fL_f}{2}\right)\norm{g[k]}^2.
\end{equation*}
Thus $0<T_sk_f<2/L_f$ yields strict descent away from the desired rigid orbit. Local infinitesimal rigidity also implies a local gradient-dominance bound $\norm g^2\ge2\mu_fV_f$ on the quotient space for some $\mu_f>0$, so
\begin{align}
 V_f[k+1]&\le\rho_fV_f[k],\nonumber\\
 \rho_f&=1-2T_sk_f\mu_f
 \left(1-\frac{T_sk_fL_f}{2}\right)<1.
 \label{eq:shape-contraction}
\end{align}
The finite rotations in \eqref{eq:xi-desired} preserve all inter-agent distances, so they do not change $V_f$. By construction, the first rotation $R(\delta_s)$ restores the anchor-vector direction to $q_F[k]$, and the second rotation applies exactly the optimized increment in \eqref{eq:orientation-increment}. The shape descent therefore does not contaminate the orientation update.

\begin{theorem}[Local convergence of the discrete tri-control law]\label{thm:closed-loop}
Suppose Assumptions~\ref{ass:info}--\ref{ass:graph} hold, the exact proximal estimator problems satisfy the uniform exact-penalty conditions of Lemma~\ref{lem:finite-prox}, $\rho_F[k]\ge\rho_{\min}>0$, and the trajectory remains in a rigidity neighborhood $\mathcal U$ in which $\nabla V_f$ is $L_f$-Lipschitz and gradient dominated. If
\begin{equation}
 0<T_sk_f<\frac{2}{L_f}
 \label{eq:shape-step-size}
\end{equation}
and $\abs{\phi[K_E]}<\pi/2$, then the exact-proximal distributed algorithm satisfies $\hat c_i[k]-\zc[k]\to0$, $\hat q_i[k]-q_F[k]\to0$ with finite-step exact consensus, and $\zc[k]-\zstar[k]\to0$, $\phi[k]\to0$, $e[k]\to0$ locally exponentially after $K_E$.
\end{theorem}

\begin{proof}
Finite-step estimator convergence follows from Lemma~\ref{lem:finite-prox}, Corollary~\ref{cor:c-r}, and the error-coordinate form of \eqref{eq:xi-prox}. After $K_E$, the geometric estimates are exact and the three optimization coordinates separate. Equations \eqref{eq:centroid-contraction}, \eqref{eq:q-contraction}, and \eqref{eq:shape-contraction} provide independent local contractions for centroid, intrinsic orientation, and shape. The horizontal-lift rotation preserves shape while eliminating the shape-induced anchor-direction change. A weighted sum of $\norm{e_c}^2$, $V_q$, and $V_f$ therefore contracts geometrically.
\end{proof}

\begin{remark}[Hypotheses in practice]\label{rem:hypotheses}
Theorem~\ref{thm:closed-loop} is local. It requires the anchor to stay away from the centroid ($\rho_F\ge\rho_{\min}$), the shape to stay in a rigidity neighborhood, and the orientation error at estimator settling to satisfy $\abs{\phi[K_E]}<\pi/2$. With the anchor at a vertex of the target polygon, the first condition holds whenever the shape is close to the target. The staged start-up and reference phase-matching of Section~\ref{sec:implementation} are practical means of meeting the other two when the robots start from an arbitrary pose: the shape is formed before tracking begins, and tracking is released toward a reference point close to the formation's actual centroid and heading.
\end{remark}

\section{Simulation, Implementation, and Experimentation}\label{sec:validation}

\subsection{Physics-based simulation with Gazebo}\label{sec:gazebo}
The Gazebo validation (Ignition Fortress) tests the deployed configuration with physically integrated motion, actuator saturation, and imperfect sensing. Each robot is modeled as a kinematic differential-drive body driven by the zero-order-hold commands $(v_i,\omega_i)$ of Section~\ref{sec:problem}. The graph, reference, and gains are those of the deployed configuration in Section~\ref{sec:testbed}. The speed limit is $u_{\max}=5$~m/s, and each measured pose passes through a fixed $5$~ms delay and small Gaussian noise that approximate the motion-capture pipeline. Because the robots start near the target formation, no staged start-up is needed.

\begin{figure}[!tbp]
 \centering
 \includegraphics[width=\columnwidth,trim=0 130 0 0,clip]{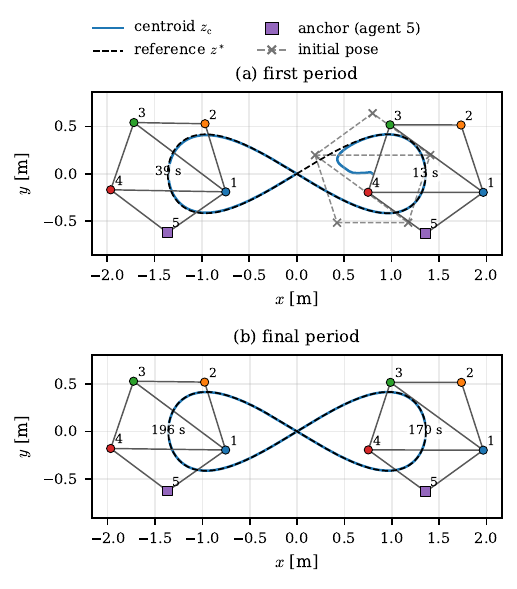}\\[-1pt]
 \includegraphics[width=\columnwidth,trim=0 0 0 150.5,clip]{fig_gazebo_xy.pdf}%
 \caption{Gazebo validation. (a) Initial and first-period formation motion. (b) Formation motion during the final reference period.}
 \label{fig:gazebo-xy}
\end{figure}

\begin{figure}[!tbp]
 \centering
 \includegraphics[width=\columnwidth]{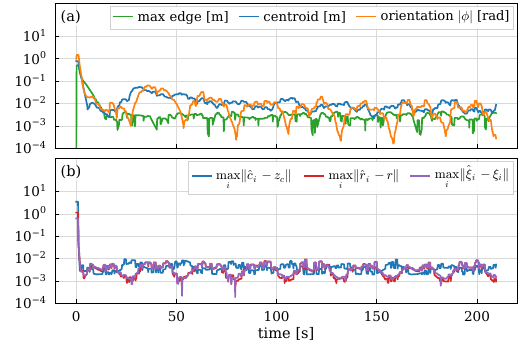}%
 \caption{Gazebo errors (1-s running maximum). (a) Shape, centroid, and orientation errors. (b) Geometric reconstruction errors.}
 \label{fig:gazebo-errors}
\end{figure}

The run covers four reference periods ($209.5$~s, $2096$ samples). The formation reaches the target pentagon during the first lobe and then repeats the figure eight with the same shape and orientation (Fig.~\ref{fig:gazebo-xy}). The edge error settles within the first seconds. The centroid and orientation errors show one excursion in the first period, reaching $5.6$~cm and $0.065$~rad near $t=32$~s, and from the second period on they stay below $1.8$~cm and $0.034$~rad without drift. In the final period, the maximum centroid and orientation errors are about $1.5$~cm and $0.015$~rad, their root-mean-square values are about $0.7$~cm and $0.005$~rad, and the maximum edge error is below $5$~mm (Table~\ref{tab:results}). The reconstruction errors fall below $1$~cm within $1$~s and stay there (Fig.~\ref{fig:gazebo-errors}). They are nonzero because the plant evolves continuously between samples, the realized increment differs slightly from the commanded one, and the sensing is delayed and noisy, whereas the estimators assume exactly realized increments (Section~\ref{sec:implementation}). Tracking nevertheless remains bounded and non-diverging for the whole run.

\subsection{Experimental platform and configuration}\label{sec:testbed}
\begin{figure}[!tbp]
 \centering
 \begin{tikzpicture}
  \node[anchor=south west, inner sep=0] (rob) at (0,0) {\includegraphics[width=0.40\columnwidth]{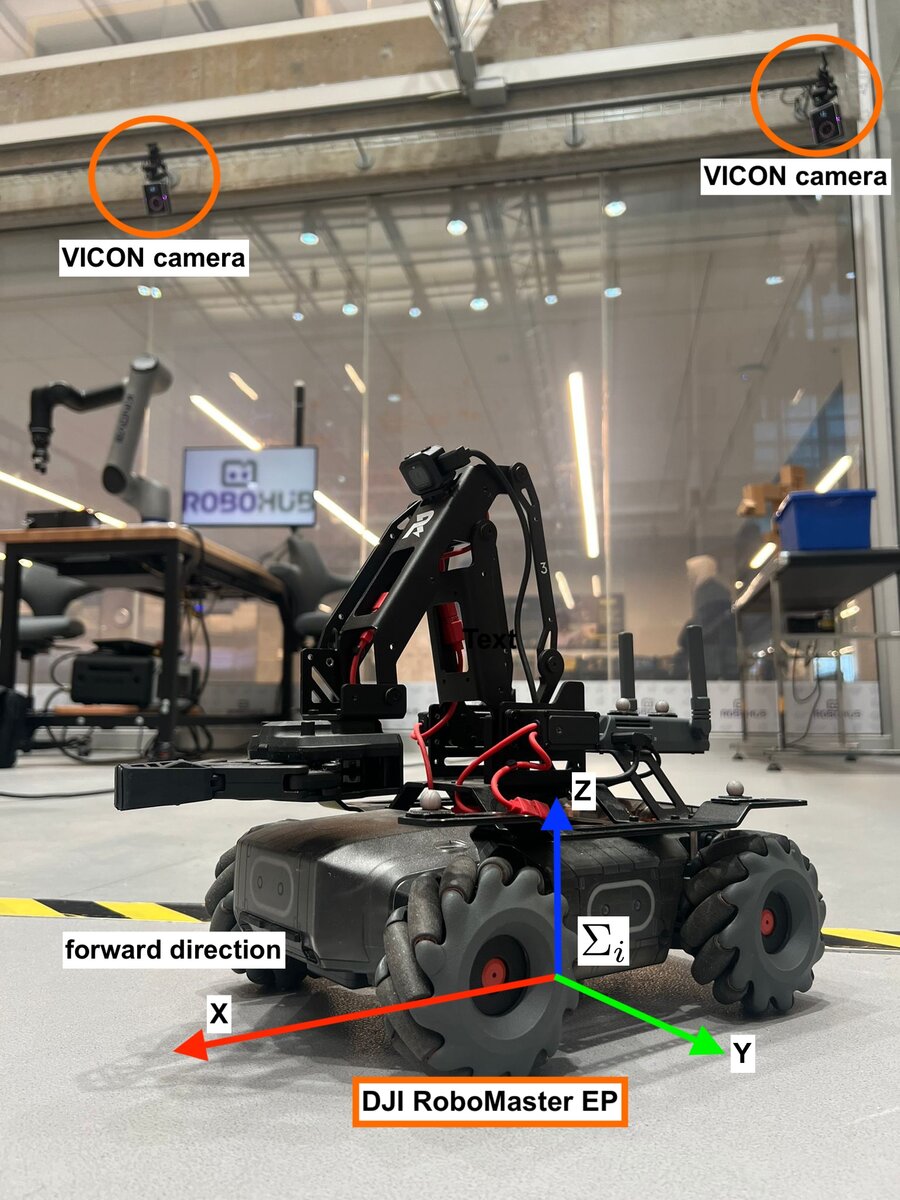}};
  \node[anchor=south west, inner sep=0] (fl) at (0.43\columnwidth,0.25) {\includegraphics[width=0.55\columnwidth]{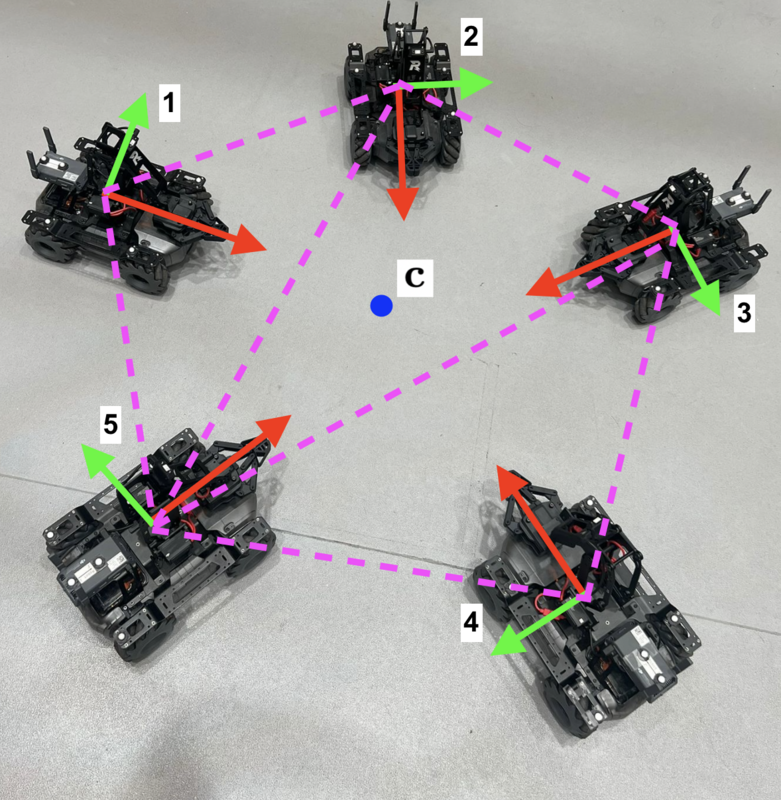}};
  \node[font=\footnotesize, anchor=north west, fill=white, inner sep=1pt] at (rob.north west) {(a)};
  \node[font=\footnotesize, anchor=north west, fill=white, inner sep=1pt] at (fl.north west) {(b)};
 \end{tikzpicture}%
 \caption{Experimental testbed. (a) DJI RoboMaster EP, body frame $\Sigma_i$, and Vicon system. (b) Five-robot formation and sensing graph.}
 \label{fig:testbed}
\end{figure}

The hardware consists of five DJI RoboMaster EP robots at the RoboHub, University of Waterloo (Fig.~\ref{fig:testbed}). The platform has mecanum wheels, but its lateral velocity is held at zero, so each robot operates as a differential-drive vehicle with inputs $(v_i,\omega_i)$. Each robot carries a body-fixed frame $\Sigma_i$ whose $x$-axis is aligned with $q_i$, and its controlled point is $z_i=p_i+\ell q_i$ with $\ell=0.22$~m. An 11-camera Vicon system measures the robot poses, and a one-time calibration per robot maps them to the kinematic frame. All robots and the workstation share one wireless network running ROS~2 Humble.

The information structure is distributed even though all computation runs on one workstation. For each robot, the workstation forms only the neighbor measurements $y^{\rm b}_{ij}=R(q_i)^{\mathsf T}(z_j-z_i)$ from the Vicon poses, which emulates on-board relative sensing, and only the anchor's absolute position is passed to the estimators. PDHG and angle-consensus messages are exchanged in-process and only along graph edges, and the resulting velocity commands are sent to the robots over the wireless network. A convex-hull safety filter, clocked independently of the controller, keeps the team inside the laboratory boundary.

The deployed graph is
\begin{equation*}
\mathcal E=\{(1,2),(1,3),(1,4),(1,5),(2,3),(3,4),(4,5)\},
\end{equation*}
with agent $5$ as the anchor, a regular-pentagon target of side length $0.75$~m and diagonal length $1.2135$~m, and $\lambda_2(L)=1.5858$. The reference centroid is the Gerono figure eight
\begin{equation*}
 \zstar(t)=c_0+R(\psi)\begin{bmatrix}a_x\sin\sigma\\
 a_y\sin\sigma\cos\sigma\end{bmatrix},
 \qquad \sigma=0.12t,
\end{equation*}
whose period is $2\pi/0.12\approx52.36$~s; $\qstar$ is its unit tangent. Gazebo uses $a_x=1.36$~m, $a_y=0.833$~m, $c_0=0$, and $\psi=0$. For the hardware run, the same curve is resized, translated, and rotated to fit inside the laboratory safety boundary, with a clearance equal to the pentagon circumradius plus the look-ahead and a margin: $a_x=1.906$~m, $a_y=1.168$~m, $c_0=[-0.038\;\;0.135]^{\mathsf T}$~m, and $\psi=90^\circ$. The deployed loop runs at $T_s=0.1$~s (10~Hz). The control gains are $k_c=0.9$, $k_q=1.5$, and $k_f=1.1$, the estimator penalties are $\alpha_c=\alpha_r=\alpha_\xi=2$ and $(\beta_c,\beta_r,\beta_\xi)=(8,18,30)$, and each sample uses $50$ PDHG rounds per estimator and $20$ angle-consensus rounds, with $\tau=\sigma=0.88/\sqrt{\lambda_{\max}(L)}$ and $\vartheta=1$. These parameters are shared by the Gazebo and hardware runs, both of which last four reference periods. In the physical run, the robots start from a distorted pentagon far from the reference, with headings between $-123^\circ$ and $49^\circ$ and axle centers spread over about $1.8\times1.9$~m.



\subsection{Implementation considerations}\label{sec:implementation}
The analysis assumes that increments are realized exactly and instantaneously, and real-time deployment exposed two effects that it does not cover.

\emph{Saturation ordering.} The estimator predictor uses the commanded $\Delta z_i[k]$ as its only motion input. If the safety filter clamps a command after that increment has been stored, the estimator propagates a motion that never happened; in testing, this mismatch made the estimator diverge within seconds. The speed limit $u_{\max}=5$~m/s is therefore applied to $u_i[k]=\Delta z_i[k]/T_s$ before $\Delta z_i[k]$ is stored for the next estimator update, and the limit is set above the normal transient peak so that the safety filter stays inactive in nominal operation.

\emph{Sensing latency.} The anchor's absolute position is used only at initialization (Section~\ref{sec:algorithm}), and the sampled estimator has no bounded restoring term, so stale relative measurements degrade the reconstruction directly. In our tests, end-to-end latency above roughly $T_s/2$ caused persistent command saturation, and latency above about $75$~ms made the estimators diverge. The deployed node therefore forms $y^{\rm b}_{ij}$ directly from the pose messages it receives instead of through an extra relay node, and it runs at 10~Hz. The measured computation time of 11--14~ms per sample shows that the inner iterations are not the bottleneck.

The physical run also starts from an arbitrary pose, which required two initialization procedures that leave the control law unchanged.

\emph{Staged start-up.} Activating \eqref{eq:discrete-tricontrol} immediately can take the system outside the local regime of Theorem~\ref{thm:closed-loop}. The commands are therefore held at zero for $10$~s while the estimators converge. Then only the shape step \eqref{eq:shape-step} is enabled for $20$~s, with $c_i^d$ and $\Delta\theta_i^d$ held at zero, and the full controller is activated at $t=30$~s.

\emph{Reference phase-matching.} The reference keeps moving during start-up, so releasing the controller at a fixed time can produce a combined centroid and orientation correction large enough to violate the local shape-step regime \eqref{eq:shape-step-size}. At mission start, we therefore select a constant reference-time offset by searching one period for the desired centroid and heading that are jointly closest to the centroid and heading of the formation at the scheduled release. The same offset is used for the rest of the run. It changes only the reference initialization, not the controller or its analysis.

\subsection{Physical Experimentation}\label{sec:physical}
The hardware run uses the same gains, estimator penalties, inner-iteration counts, and sampling rate as the Gazebo run, but it starts from the arbitrary initial poses described in Section~\ref{sec:testbed}, so the procedures above are applied: zero commands for $10$~s, shape-only control from $10$ to $30$~s, and full tracking from $t=30$~s. At release, the phase-matched reference is $0.45$~m from the formation centroid and has nearly the same heading.

\begin{figure}[!tbp]
 \centering
 \includegraphics[width=\columnwidth]{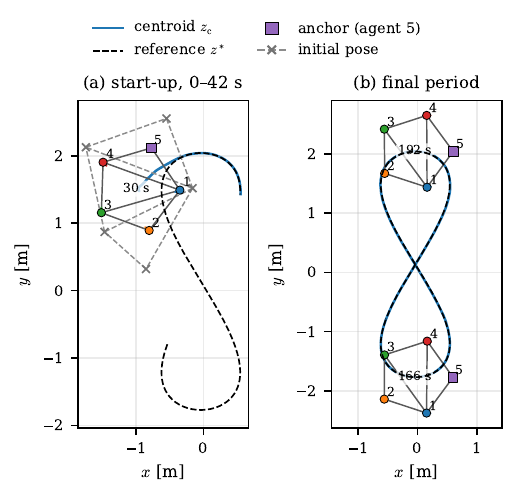}%
 \caption{Physical experiment. (a) Formation initialization and tracking release. (b) Formation motion during the final reference period.}
 \label{fig:physical-xy}
\end{figure}

\begin{figure}[!tbp]
 \centering
 \includegraphics[width=\columnwidth]{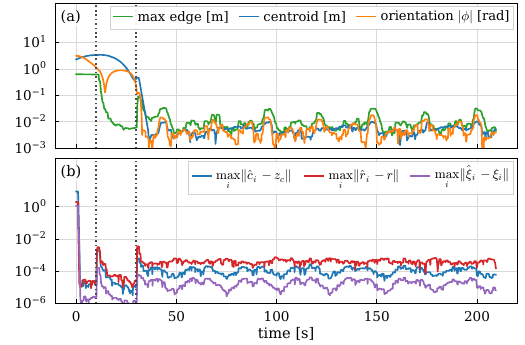}%
 \caption{Physical experiment errors (1-s running maximum). (a) Shape, centroid, and orientation errors. (b) Geometric reconstruction errors.}
 \label{fig:physical-errors}
\end{figure}

During start-up, the scattered robots form the pentagon in the shape-only stage, and once full control is released, the centroid joins the reference along a short path (Fig.~\ref{fig:physical-xy}). The edge error drops by two orders of magnitude between $t=10$ and $30$~s. After release, the orientation error falls below $0.05$~rad within $0.1$~s, and the centroid error falls below $5$~cm within $2.7$~s (Fig.~\ref{fig:physical-errors}). All three errors then stay bounded without drift, and in the final period the formation tracks the figure eight steadily while keeping its shape and orientation.

Table~\ref{tab:results} compares the two validations. On hardware, the final-period centroid and orientation errors are at most $1.04$~cm and $0.018$~rad, with root-mean-square values of $0.48$~cm and $0.005$~rad. These errors are of the same order as in Gazebo, even though the robots started from an arbitrary pose. The shape errors are larger on hardware (maximum edge error $3.2$~cm, root-mean-square $1.0$~cm). This is consistent with effects that the kinematic simulation does not include: slip of the mecanum wheels under differential-drive operation, residual calibration error between the Vicon frame and the kinematic frame, and network and sensing delays. The hardware reconstruction errors are smaller than in Gazebo because the estimates and the values they are compared with both come from the same motion-capture poses. These errors therefore measure how consistent the reconstruction is, not the absolute accuracy of the Vicon system. To the authors' knowledge, this is the first validation of this algorithm on physical hardware.

\begin{table}[!tbp]
 \centering
 \caption{Final-period errors: maximum (root mean square)}
 \label{tab:results}
 \footnotesize
 \setlength{\tabcolsep}{2.5pt}
 \begin{tabular}{@{}lcc@{}}
  \toprule
  Metric & Gazebo & Physical\\
  \midrule
  Max edge error [mm] & 4.5 (2.1) & 32.1 (9.7)\\
  Centroid tracking [mm] & 14.6 (6.8) & 10.4 (4.8)\\
  Orientation $\abs\phi$ [mrad] & 14.8 (5.3) & 18.1 (5.0)\\
  \midrule
  Centroid estimation [mm] & 9.3 (2.5) & 0.23 (0.07)\\
  Anchor-vector estimation [mm] & 8.5 (2.3) & 0.67 (0.24)\\
  Centered-coord.\ estimation [mm] & 8.7 (2.4) & 0.04 (0.01)\\
  Centroid consensus [mm] & 1.0 (0.18) & 0.45 (0.14)\\
  \bottomrule
 \end{tabular}\\[2pt]
 \parbox{\columnwidth}{\scriptsize Hardware estimation errors are measured against the motion-capture poses from which the measurements are generated.}
\end{table}

\section{Conclusion}\label{sec:conclusion}
We derived a distributed estimator--controller for rigid differential-drive formations from an optimization-based formulation. The centroid, the anchor vector, and the centered coordinates are reconstructed directly from one absolute-position anchor and neighbor relative-position measurements, without reconstructing the absolute position of any individual robot. Exact graph-proximal updates complete this reconstruction in finitely many steps, and a finite number of neighbor-only primal--dual rounds leaves a bounded estimation error. The controller regulates the centroid, the intrinsic orientation, and the distance-based shape separately, and a finite rotation removes the orientation drift caused by the shape step, which gives local exponential convergence once the reconstruction is exact.

We validated the method in ROS~2/Gazebo with imperfect sensing and actuation and in experiments with five physical robots. On hardware, the final-period centroid and orientation errors stayed below $1.1$~cm and $0.019$~rad, and the formation was maintained after starting from an arbitrary initial pose. The experiments also showed the importance of three practical measures: feeding the applied increments to the estimator, accounting for sensing latency, and using staged start-up with reference phase-matching.

Future work will extend the framework to Ackermann-steered vehicles. There, the locally unconstrained planar-increment realization will be replaced by a curvature-constrained motion model that accommodates the minimum turning radius and steering limits while retaining centroid tracking, formation-orientation tracking, and rigid-shape regulation.

\bibliographystyle{IEEEtran}
\bibliography{references}

@article{AndersonEtAl2008,
  author  = {B. D. O. Anderson and C. Yu and B. Fidan and J. M. Hendrickx},
  title   = {Rigid graph control architectures for autonomous formations},
  journal = {IEEE Control Systems Magazine},
  volume  = {28},
  number  = {6},
  pages   = {48--63},
  year    = {2008}
}

@article{KrickEtAl2009,
  author  = {L. Krick and M. E. Broucke and B. A. Francis},
  title   = {Stabilisation of infinitesimally rigid formations of multi-robot networks},
  journal = {International Journal of Control},
  volume  = {82},
  number  = {3},
  pages   = {423--439},
  year    = {2009}
}

@article{deMarinaEtAl2016,
  author  = {H. {Garc{\'i}a de Marina} and B. Jayawardhana and M. Cao},
  title   = {Distributed rotational and translational maneuvering of rigid formations and their applications},
  journal = {IEEE Transactions on Robotics},
  volume  = {32},
  number  = {3},
  pages   = {684--697},
  year    = {2016}
}

@article{YangEtAl2018,
  author  = {Q. Yang and M. Cao and H. Garc{\'i}a de Marina and H. Fang and J. Chen},
  title   = {Distributed formation tracking using local coordinate systems},
  journal = {Systems \& Control Letters},
  volume  = {111},
  pages   = {70--78},
  year    = {2018}
}

@article{SunEtAl2017,
  author  = {Z. Sun and M.-C. Park and B. D. O. Anderson and H.-S. Ahn},
  title   = {Distributed stabilization control of rigid formations with prescribed orientation},
  journal = {Automatica},
  volume  = {78},
  pages   = {250--257},
  year    = {2017}
}

@article{ChoopojcharoenEtAl2025,
  author  = {T. Choopojcharoen and E. Selim and B. Fidan},
  title   = {Distributed simultaneous centroid estimation and formation tracking control using relative position measurement},
  journal = {IEEE Control Systems Letters},
  volume  = {9},
  pages   = {1297--1302},
  year    = {2025}
}

@inproceedings{HerguedasEtAl2022,
  author    = {R. Herguedas and M. Aranda and G. L{\'o}pez-Nicol{\'a}s and C. Sag{\"u}{\'e}s and Y. Mezouar},
  title     = {Multirobot control with double-integrator dynamics and control barrier functions for deformable object transport},
  booktitle = {Proc. IEEE Int. Conf. Robotics and Automation},
  pages     = {1485--1491},
  year      = {2022}
}

@inproceedings{FreemanYangLynch2006,
  author    = {R. A. Freeman and P. Yang and K. M. Lynch},
  title     = {Stability and convergence properties of dynamic average consensus estimators},
  booktitle = {Proc. 45th IEEE Conf. Decision Control},
  pages     = {338--343},
  year      = {2006}
}

@article{KiaEtAl2019,
  author  = {S. S. Kia and B. {Van Scoy} and J. Cort{\'e}s and R. A. Freeman and K. M. Lynch and S. Mart{\'i}nez},
  title   = {Tutorial on dynamic average consensus: The problem, its applications, and the algorithms},
  journal = {IEEE Control Systems Magazine},
  volume  = {39},
  number  = {3},
  pages   = {40--72},
  year    = {2019}
}

@article{ChenCaoRen2012,
  author  = {F. Chen and Y. Cao and W. Ren},
  title   = {Distributed average tracking of multiple time-varying reference signals with bounded derivatives},
  journal = {IEEE Transactions on Automatic Control},
  volume  = {57},
  number  = {12},
  pages   = {3169--3174},
  year    = {2012}
}

@article{CaoRenMeng2010,
  author  = {Y. Cao and W. Ren and Z. Meng},
  title   = {Decentralized finite-time sliding mode estimators and their applications in decentralized finite-time formation tracking},
  journal = {Systems \& Control Letters},
  volume  = {59},
  number  = {9},
  pages   = {522--529},
  year    = {2010}
}

@article{ZhuMartinez2010,
  author  = {M. Zhu and S. Mart{\'i}nez},
  title   = {Discrete-time dynamic average consensus},
  journal = {Automatica},
  volume  = {46},
  number  = {2},
  pages   = {322--329},
  year    = {2010}
}

@article{MontijanoEtAl2014,
  author  = {E. Montijano and J. I. Montijano and C. Sag{\"u}{\'e}s and S. Mart{\'i}nez},
  title   = {Robust discrete time dynamic average consensus},
  journal = {Automatica},
  volume  = {50},
  number  = {12},
  pages   = {3131--3138},
  year    = {2014}
}

@article{FranceschelliGasparri2019,
  author  = {M. Franceschelli and A. Gasparri},
  title   = {Multi-stage discrete time and randomized dynamic average consensus},
  journal = {Automatica},
  volume  = {99},
  pages   = {69--81},
  year    = {2019}
}

@article{AraguesEtAl2012,
  author  = {R. Arag{\"u}{\'e}s and L. Carlone and C. Sag{\"u}{\'e}s and G. Calafiore},
  title   = {Distributed centroid estimation from noisy relative measurements},
  journal = {Systems \& Control Letters},
  volume  = {61},
  number  = {7},
  pages   = {773--779},
  year    = {2012}
}

@article{FranceschelliGasparri2014,
  author  = {M. Franceschelli and A. Gasparri},
  title   = {Gossip-based centroid and common reference frame estimation in multiagent systems},
  journal = {IEEE Transactions on Robotics},
  volume  = {30},
  number  = {2},
  pages   = {524--531},
  year    = {2014}
}

@article{ZhaoEtAl2018,
  author  = {S. Zhao and D. V. Dimarogonas and Z. Sun and D. Bauso},
  title   = {A general approach to coordination control of mobile agents with motion constraints},
  journal = {IEEE Transactions on Automatic Control},
  volume  = {63},
  number  = {5},
  pages   = {1509--1516},
  year    = {2018}
}

@article{KhaledyanEtAl2020,
  author  = {M. Khaledyan and T. Liu and V. Fernandez-Kim and M. de Queiroz},
  title   = {Flocking and target interception control for formations of nonholonomic kinematic agents},
  journal = {IEEE Transactions on Control Systems Technology},
  volume  = {28},
  number  = {4},
  pages   = {1603--1610},
  year    = {2020}
}

@article{WilsonEtAl2020,
  author  = {S. Wilson and P. Glotfelter and L. Wang and S. Mayya and G. Notomista and M. Mote and M. Egerstedt},
  title   = {The Robotarium: Globally impactful opportunities, challenges, and lessons learned in remote-access, distributed control of multirobot systems},
  journal = {IEEE Control Systems Magazine},
  volume  = {40},
  number  = {1},
  pages   = {26--44},
  year    = {2020}
}

@article{Cortes2008,
  author  = {J. Cort{\'e}s},
  title   = {Discontinuous dynamical systems: A tutorial on solutions, nonsmooth analysis, and stability},
  journal = {IEEE Control Systems Magazine},
  volume  = {28},
  number  = {3},
  pages   = {36--73},
  year    = {2008}
}

@article{NedicOzdaglar2009,
  author  = {A. Nedi{\'c} and A. Ozdaglar},
  title   = {Distributed subgradient methods for multi-agent optimization},
  journal = {IEEE Transactions on Automatic Control},
  volume  = {54},
  number  = {1},
  pages   = {48--61},
  year    = {2009}
}

@article{BoydEtAl2011,
  author  = {S. Boyd and N. Parikh and E. Chu and B. Peleato and J. Eckstein},
  title   = {Distributed optimization and statistical learning via the alternating direction method of multipliers},
  journal = {Foundations and Trends in Machine Learning},
  volume  = {3},
  number  = {1},
  pages   = {1--122},
  year    = {2011}
}

@article{ParikhBoyd2014,
  author  = {N. Parikh and S. Boyd},
  title   = {Proximal algorithms},
  journal = {Foundations and Trends in Optimization},
  volume  = {1},
  number  = {3},
  pages   = {127--239},
  year    = {2014}
}

@article{ChambollePock2011,
  author  = {A. Chambolle and T. Pock},
  title   = {A first-order primal-dual algorithm for convex problems with applications to imaging},
  journal = {Journal of Mathematical Imaging and Vision},
  volume  = {40},
  number  = {1},
  pages   = {120--145},
  year    = {2011}
}

@article{BurkeFerris1993,
  author  = {J. V. Burke and M. C. Ferris},
  title   = {Weak sharp minima in mathematical programming},
  journal = {SIAM Journal on Control and Optimization},
  volume  = {31},
  number  = {5},
  pages   = {1340--1359},
  year    = {1993}
}

@article{MatsushitaXu2012,
  author  = {S.-y. Matsushita and L. Xu},
  title   = {Finite termination of the proximal point algorithm in Banach spaces},
  journal = {Journal of Mathematical Analysis and Applications},
  volume  = {387},
  number  = {2},
  pages   = {765--769},
  year    = {2012}
}

@article{LiuEtAl2021,
  author  = {D. Liu and W. Bao and X. Zhu and B. Fei and T. Men and Z. Xiao},
  title   = {Cooperative path optimization for multiple UAVs surveillance in uncertain environment},
  journal = {IEEE Internet of Things Journal},
  volume  = {9},
  number  = {13},
  pages   = {10676--10692},
  year    = {2022}
}

@article{CuiEtAl2024,
  author  = {Y. Cui and Y. Huang and Q. Zhao and X. Yu and T. Huang},
  title   = {Resilient time-varying formation-tracking of multiagent systems against hybrid attacks with applications to spacecraft formation},
  journal = {IEEE Transactions on Cybernetics},
  volume  = {55},
  number  = {2},
  pages   = {625--637},
  year    = {2025}
}

@article{ReckerEtAl2021,
  author  = {T. Recker and M. Heinrich and A. Raatz},
  title   = {A comparison of different approaches for formation control of nonholonomic mobile robots regarding object transport},
  journal = {Procedia CIRP},
  volume  = {96},
  pages   = {248--253},
  year    = {2021}
}

@article{WuEtAl2024,
  author  = {Q. Wu and X. Wang and X. Qiu},
  title   = {Embedded technique-based formation control of multiple wheeled mobile robots with application to cooperative transportation},
  journal = {Control Engineering Practice},
  volume  = {150},
  pages   = {106002},
  year    = {2024}
}

@article{RenEtAl2020,
  author  = {Y. Ren and S. Sosnowski and S. Hirche},
  title   = {Fully distributed cooperation for networked uncertain mobile manipulators},
  journal = {IEEE Transactions on Robotics},
  volume  = {36},
  number  = {4},
  pages   = {984--1003},
  year    = {2020}
}

@article{HuEtAl2021,
  author  = {J. Hu and P. Bhowmick and I. Jang and F. Arvin and A. Lanzon},
  title   = {A decentralized cluster formation containment framework for multirobot systems},
  journal = {IEEE Transactions on Robotics},
  volume  = {37},
  number  = {6},
  pages   = {1936--1955},
  year    = {2021}
}

@article{ZhangEtAl2025,
  author  = {X. Zhang and Q. Yang and F. Xiao and H. Fang and J. Chen},
  title   = {Linear formation control of multi-agent systems},
  journal = {Automatica},
  volume  = {171},
  pages   = {111935},
  year    = {2025}
}

@article{PengEtAl2025,
  author  = {C. Peng and J. Huang},
  title   = {Angle-constrained distributed formation control with disturbance rejection},
  journal = {Automatica},
  volume  = {177},
  pages   = {112314},
  year    = {2025}
}

@inproceedings{HabibiEtAl2015,
  author    = {G. Habibi and Z. Kingston and W. Xie and M. Jellins and J. McLurkin},
  title     = {Distributed centroid estimation and motion controllers for collective transport by multi-robot systems},
  booktitle = {Proc. IEEE Int. Conf. Robotics and Automation},
  pages     = {1282--1288},
  year      = {2015}
}

@article{SunEtAl2022,
  author  = {Y. Sun and Y. Zou and X. He and Q. Fu and J. Wu},
  title   = {Distributed formation centroid tracking control of clustered rotorcraft},
  journal = {IEEE Transactions on Aerospace and Electronic Systems},
  volume  = {59},
  number  = {2},
  pages   = {1867--1878},
  year    = {2023}
}

@article{GuoEtAl2020,
  author  = {M. Guo and B. Jayawardhana and J. G. Lee and H. Shim},
  title   = {Simultaneous distributed localization, mapping and formation control of mobile robots based on local relative measurements},
  journal = {IFAC-PapersOnLine},
  volume  = {53},
  number  = {2},
  pages   = {9614--9620},
  year    = {2020}
}

@article{FangEtAl2025,
  author  = {X. Fang and L. Xie and D. V. Dimarogonas},
  title   = {Simultaneous distributed localization and formation tracking control via matrix-weighted position constraints},
  journal = {Automatica},
  volume  = {175},
  pages   = {112188},
  year    = {2025}
}

@inproceedings{TsoukalasEtAl2018,
  author    = {A. Tsoukalas and A. Tzes and F. Khorrami},
  title     = {Relative pose estimation of unmanned aerial systems},
  booktitle = {Proc. Mediterranean Conf. Control and Automation},
  pages     = {155--160},
  year      = {2018}
}

@article{WangEtAl2014,
  author  = {X. Wang and Y. A. {\c S}ekercio{\u g}lu and T. Drummond},
  title   = {Vision-based cooperative pose estimation for localization in multi-robot systems equipped with RGB-D cameras},
  journal = {Robotics},
  volume  = {4},
  number  = {1},
  pages   = {1--22},
  year    = {2015}
}

\end{document}